\documentclass[unstructuredabstract]{aa}

\usepackage{graphicx}
\usepackage{txfonts}
\usepackage{siunitx}
\usepackage[colorlinks,citecolor=blue,urlcolor=blue,filecolor=blue,linkcolor=blue]{hyperref}
\usepackage{textgreek}

\begin{document} 
   \title{Pulsars identified in the LOFAR Two-metre Sky Survey at 144\,MHz}

\author{
  G.\,A.\,C.\,Rijkers\inst{\ref{leiden}}
  \and
  C.\,G.\,Bassa\inst{\ref{astron}}
  \and
  J.\,R.\,Callingham\inst{\ref{astron},\ref{api}}
  \and
  T.\,Shimwell\inst{\ref{astron},\ref{leiden}}
}

\institute{
  Leiden Observatory, Leiden University, P.O. Box 9500, 2300 RA Leiden, The Netherlands\\ \email{rijkers@strw.leidenuniv.nl}
  \label{leiden}
  \and
  ASTRON, Netherlands Institute for Radio Astronomy, Oude Hoogeveensedijk 4, 7991 PD Dwingeloo, The Netherlands
  \label{astron}
  \and
  Anton Pannekoek Institute for Astronomy, University of Amsterdam, Science Park 904, 1098 XH Amsterdam, The Netherlands\label{api}
}

\date{Received September 29, 2025; Accepted December 8, 2025}

\abstract{We present the astrometric identification of 80 known radio pulsars as unresolved continuum sources detected at 144\,MHz in the second data release of the LOFAR Two-metre Sky Survey (LoTSS DR2), which covers 27\% of the Northern hemisphere. These identifications represent the majority ($\geq 86\%$) of radio pulsars present in the LoTSS DR2 footprint and provide independent celestial positions and flux densities at 144\,MHz. We compare LoTSS flux densities with literature values from various image and time-domain observations and find good agreement for all but two pulsars. We attribute these flux density deviations to intrinsic pulsar properties (nulling and off-pulse emission). We investigate criteria to select promising pulsar candidates using data from the upcoming LoTSS release of the entire Northern sky ($\delta>0$\textdegree), as well as the LOFAR LBA Sky Survey (LoLSS) at 54\,MHz (covering $\delta>24$\textdegree). Of the 80 detections, 35 (44\%) were blindly redetected based on their linear or circular polarization. Therefore we conclude that candidate selection based on polarization properties is a promising approach. Candidate selection can be supplemented with spectral indices via cross-matching to LoLSS sources at 54\,MHz, as the high sensitivity of LoTSS is not matched by image-domain surveys at higher frequencies.} 


\maketitle

\nolinenumbers
\section{Introduction}
By nature of their regular pulsations, radio pulsars offer unique probes into various astrophysical processes through the method of pulsar timing \citep[see review by][]{2003LRR.....6....5S}. These processes range from measuring neutron star properties \citep[e.g.][]{2021ApJ...915L..12F,2024MNRAS.530.1581K}, understanding the properties of the interstellar medium \citep[e.g][]{2019MNRAS.484.3646S,2023MNRAS.518.1086M}, constraining nano-Hertz gravitational waves \citep[e.g.][]{2023ApJ...951L..11A,2023A&A...678A..50E,2023ApJ...951L...6R,2023RAA....23g5024X} and characterizing the Galactic population of pulsars, and with it, that of neutron stars \citep[e.g.][]{2006ApJ...643..332F,2024ApJ...968...16G}. These results are predicated on the necessity to resolve the pulsar pulse profile, hence high-time resolution radio observations are required to resolve pulsations for even the fastest spinning radio pulsars (which have spin periods as fast as to $P=1.4$\,ms; \citealt{2006Sci...311.1901H,2017ApJ...846L..20B}).

Interferometric observations using aperture synthesis \citep{1958MNRAS.118..276J,1960MNRAS.120..220R} generally trade time resolution for spatial resolution and hence fully integrate over the pulse profile of pulsars. These phase averaged observations provide independent measurements of pulsar flux densities, polarization properties and celestial positions, and are not biased by temporal smearing of the pulse profile by dispersion and scatter broadening, which are strongly dependent on the pulsar spin period and the pulsar dispersion measure \citep{2003ApJ...596.1142C}, as well as any temporal smearing due to unresolved binary motion \citep{2002AJ....124.1788R}. Additionally, the flux scale in imaging observations is better defined than that of high time resolution single dish observations of beamformed observations with multi-element telescopes, which need to rely on accurate knowledge of instrumental parameters through the radiometer equation \citep[e.g.][]{2004hpa..book.....L}. As a result, wide-area imaging surveys provide uniform flux density measurements for large fractions of the pulsar population.

Conversely, radio pulsars typically have steep radio spectra ($S_\nu\propto\nu^\alpha$ with $\alpha=-1.6\pm1.0$; \citealt{2018MNRAS.473.4436J}), and significant linear or circular polarization fractions \citep{2023MNRAS.520.4582P} which allows for identification of radio pulsar candidates from imaging observations. This point is perhaps best illustrated by the discovery of the first radio millisecond pulsar PSR\,B1937+21 ($P=1.6$\,ms) by \citet{1982Natur.300..615B}, which was found by targeting the steep spectrum, strongly scintillating and linearly polarized point source 4C21.53\footnote{The history of this discovery has recently been recollected by \citet{2024JAHH...27..453R,2024JAHH...27..465D} and \citet{2024JAHH...27..482K}.}. 

In this paper we search for known radio pulsars that are present in the LOFAR Two-metre Sky Survey (LoTSS; \citealt{Shimwell+17}) as continuum sources. LoTSS is an ongoing radio continuum survey of the Northern hemisphere at frequencies between 120 and 168\,MHz with LOFAR \citep{vanHaarlem+13}. The survey reaches sensitivities of $100$\,\textmu Jy\,beam$^{-1}$ at a resolution of $6\arcsec$. The majority of pointings have an integration time of at least 8 hours. The latest LoTSS data release (DR2; \citealt{Shimwell+22}) provides total intensity images and a catalog containing more than 4 million sources for 27\% of the Northern hemisphere, mostly at high Galactic latitudes ($|b|>15$\degr). 

In Sect.\,\ref{sec:crossmatching} we cross match the LoTSS DR2 catalog to astrometric positions of radio pulsars. The resulting detections and non-detections are discussed in Sect.\,\ref{sec:results}, which provides flux density measurements, polarization properties and an investigation of globular cluster environments. We discuss and conclude in Sect.\,\ref{sec:discussion_and_conclusions}, where we focus on the number of radio pulsars detected in LoTSS, compare flux density measurements from image and time-domain methods, discuss selection criteria for identifying pulsar candidates, and present an outlook for upcoming LOFAR image domain data releases.

\section{Catalog cross-matching}\label{sec:crossmatching}

\begin{figure*}
\centering
   \includegraphics[width=\textwidth]{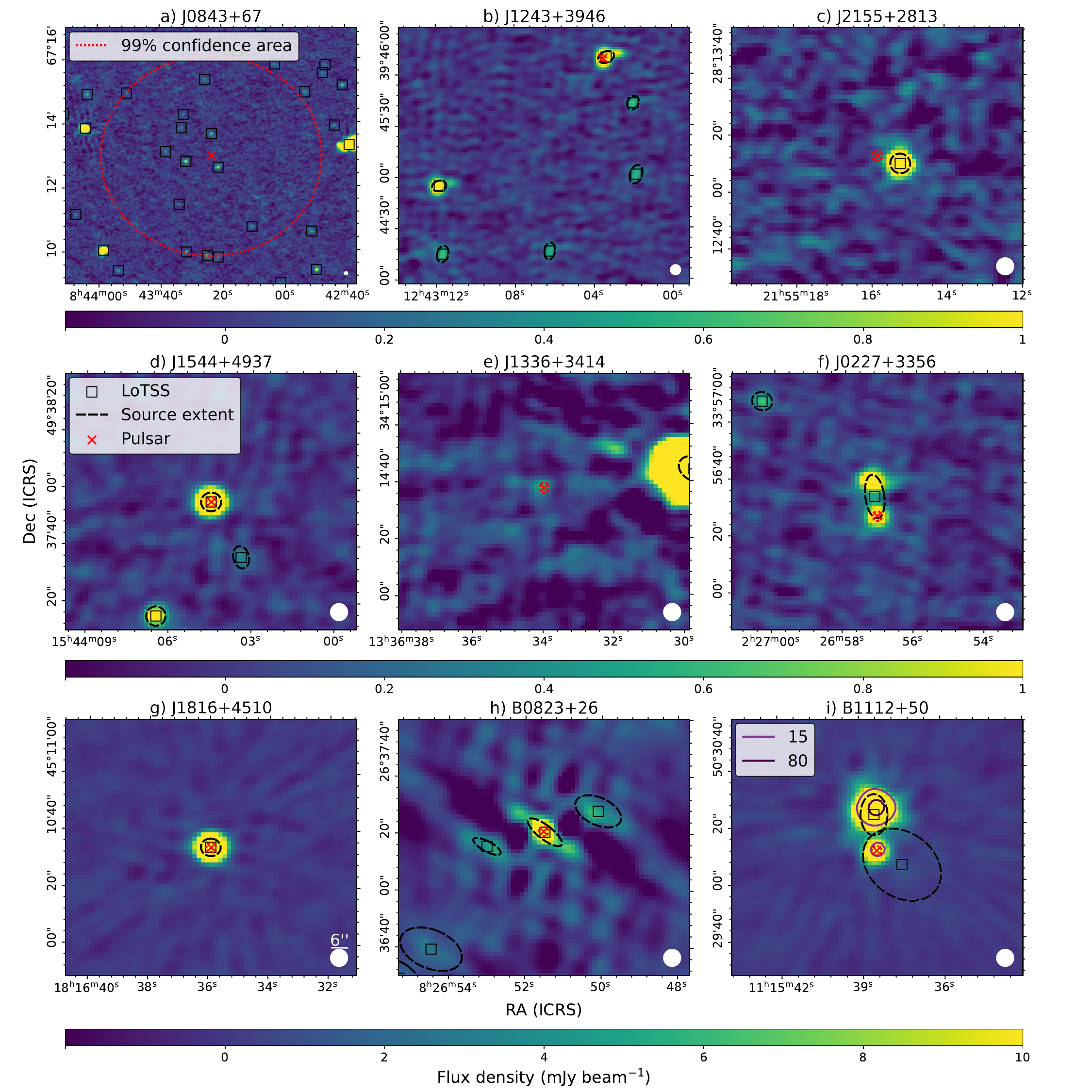}
   \caption{LoTSS DR2 images in Stokes I for nine example pulsars. All images have a field-of-view of $90\arcsec\,\times\,90\arcsec$, except for panel a) which has $8\arcmin\,\times\,8\arcmin$ and panel b) which has $2\farcm5 \times 2\farcm5$. Positions of LoTSS sources are indicated with a black square, while positions from the pulsar catalog are denoted with the red cross. The extent of LoTSS sources is shown with the black dashed ellipses, depicting the major- and minor axes resulting from fitting to the LoTSS sources \citep{Shimwell+22}. The FWHM of the LoTSS beam is shown in the bottom-right corners. Flux density contours in mJy\,beam$^{-1}$ are plotted for panel i) to highlight the structure within bright sources. Panel d) and g) show example pulsars with accurate position measurements, leading to a unique match to a LoTSS source, while panel a) shows an example where the positional uncertainty of the pulsar is large and several LoTSS sources coincide with the 99\% confidence error region (denoted by the red dotted ellipse).
   In panel b) a field of LoTSS sources are shown, where the Northern-most source is matched to a pulsar. All visible components show source extension to the East, indicating an imaging problem.
   Pulsar\,J1336+3414 is dimly visible as a point source in panel e), and due to its low brightness and/or the proximity of the Eastern bright source it is a $\sim4\sigma$ detection that is not catalogued in LoTSS DR2. We estimate its flux density directly from the image.
   Panel c) shows a pulsar with a separation of $8\farcs4$ to a LoTSS source, but does not contain it within its 99\% uncertainty ellipse. In panel f) a pulsar is matched to a LoTSS source that was classified as a single extended object, while it in fact consists of two unrelated objects, one of which is the pulsar counterpart. Finally, panel i) shows a compact radio source that is present at the position of PSR\,B1112+50, but due to the proximity of a brighter radio source, it is catalogued in LoTSS as an extended source with an offset position.}
   \label{fig:pulsar_examples}
\end{figure*}

We use the ATNF pulsar catalog (PSRCAT; \citealt{Pulsar_catalog})\footnote{\url{www.atnf.csiro.au/research/pulsar/psrcat/}}, which collates pulsar properties from the literature. We used version v2.2.0 which contains properties of 3630 pulsars, of which the majority are spin-powered radio pulsars, though the catalog also includes X-ray and gamma-ray emitting pulsars and magnetically powered magnetars. The pulsar catalog provides rotational, astrometric properties, as well as binary parameters for those pulsars in binaries.

We amend the pulsar catalog with information from recent publications. These include timing solutions for 16 pulsars discovered with LOFAR from \citet{2023A&A...669A.160V}, four Fermi $\gamma$-ray pulsars (three by \citealt{2022ApJS..260...53A}, one by \citealt{2023ApJ...958..191S}), one pulsar discovered as part of the Green Bank North Celestial Cap (GBNCC) survey \citep{2023ApJ...956...40F}, one CHIME pulsar \citep{Dong+2023} and one pulsar discovered with the Large Phased Array telescope by \citet{2017ARep...61..848T}. We also use LOFAR timing solutions for six pulsars by \citet{McKenna+2024} and \citet{2024PhDT........40M}. For PSR\,J1641+3627F in the globular cluster M\,13, the updated values from the FAST timing solution \citep{2020ApJ...892...43W} did not contain an uncertainty value in declination, and we assume it to be in the last significant digit; $\sigma_\delta=0\farcs0010$. For PSR\,J2212+2450 we use the recent timing position from \citet{2023ApJ...956..132L}. 

Proper motion is accounted for when available in PSRCAT, using the reported position epoch in PSRCAT and the observation date of the LoTSS observation. In the case where a pulsar overlaps with more than one LoTSS observation, we use the observation whose pointing is closest to the location of the pulsar. For those pulsars that have a known proper motion, it is only applied when there is also a known uncertainty on the proper motion. In general, the positional shifts due to proper motion were negligible ($0\farcs05$ or less), except for PSRs\,B1112+50 at $1\farcs14$, B1508+55 at $1\farcs28$ and J1605+3249 at $2\farcs43$, as these pulsars have high proper motion and/or positional epochs from the early 2000s.

We find that 117 pulsars in the pulsar catalog are located within the sky area surveyed in LoTSS DR2. Positional uncertainties are not available in PSRCAT for 10 pulsars, and for 13 other pulsars the positional uncertainties exceed $20\arcsec$ in one or both coordinates, precluding unique identification of pulsar counterparts. These pulsars are removed from our sample, and the total number of pulsars with reliable information that we can cross-match to LoTSS sources is 95.

For these 95 pulsars, we use the positional information to compute 99\% confidence uncertainty regions to identify counterparts in the LoTSS DR2 catalog and images. Here, the positional uncertainties in right ascension and declination of pulsars are summed in quadrature with the positional uncertainty in the frame tie of LoTSS to the ICRS. The frame tie is estimated at $0\farcs20$ in both right ascension and declination based on optical-radio matches \citep{Shimwell+22}. The typical value for the positional uncertainty of LoTSS sources is also added in the quadratic sum, for which we take the median positional uncertainty of $\sim0\farcs50$ in either coordinate for LoTSS sources with fluxes between 0.1 and 10\,mJy.

The 99\% uncertainty regions are a factor $[-2\log_e(1-0.99)]^{1/2}=3.035$ larger than the total $1\sigma$ uncertainties in either coordinate. We consider LoTSS sources coincident within the 99\% confidence pulsar uncertainty regions as counterparts. 

\section{Results}\label{sec:results}

\subsection{Detections} \label{subsec: Detections}
For the 95 pulsars with accurate positional information, we find that 80 pulsars have a LoTSS counterpart. Among those pulsars, 76 have a LoTSS source coincident with the 99\% confidence uncertainty regions. Visual inspection of the LoTSS images of all 95 pulsars reveal four further matches which are discussed below. The properties of the radio pulsars and their associated LoTSS sources are given in Table\,\ref{tab:detections}, and pulsar positions are overlaid on LoTSS continuum images for 9 pulsars in Fig.\,\ref{fig:pulsar_examples}. 

Given the LoTSS source density of 0.22 sources per square arcminute, we estimate the number of unrelated LoTSS sources coincident with one of the 99\% confidence uncertainty regions of all pulsars (covering 2.64 square arcminutes in total) at 0.6, suggesting that on the order of one pulsar may be associated with a LoTSS source due to random chance. Excluding the four pulsars matched by visual inspection, we find that the average offset between the pulsar positions and the LoTSS source positions is consistent within the uncertainties ($\Delta \alpha = 0\farcs19 \pm 0\farcs 94$ and $\Delta \delta = -0\farcs02 \pm 0\farcs94$). The cumulative density function (CDF) positional offsets we measure correspond to a Rayleigh distribution with a scale parameter of $\sigma = 0.61$. A value near unity is expected for $\sigma$, which shows that the positional uncertainties are likely overestimated. Neglecting the uncertainties in the frame-tie, which are systematic and not random, yields $\sigma=1.30$, while decreasing the frame-tie uncertainty to $0\farcs05$ in both coordinates results in $\sigma=1.05$. This leads us to conclude that our sample of matches is statistically complete. 

Four sources were visually matched to pulsars.

\emph{B1112+50:} Due to its location in the wings of the point-spread function of a brighter source (ILT\,J111538.50+503023.8, $\mathrm{S_{144}}=233$\,mJy) located $13\arcsec$ North of the pulsar position, PSR\,B1112+50 is detected in LoTSS imaging, but catalogued as an extended and offset source. The proximity of the pulsar and the brighter source appears to have impacted the LoTSS source finding algorithm, as the source positions and source extent are deviating from the peaks in the image (Fig.\,\ref{fig:pulsar_examples}i). We note that \citet{O'Sullivan+2023} matched PSR\,B1112+50 with the brighter source ILT\,J111538.50+503023.8. This mis-identification was also made by \citet{Frail+2016} due to the lower angular resolution of the TGSS survey, as they report a flux density (198\,mJy), far higher than the peak flux in LoTSS ($\mathrm{S_{144}}=21$\,mJy) and other literature. Based on NVSS observations at 1.4\,GHz, \citet{Kaplan1998} mentions that the matching of PSR\,B1112+50 was confused by a source $12\arcsec$ away, which is consistent with our observations.

\emph{J0227+3356:} Two point sources of similar brightness have been identified as a single extended source in the LoTSS catalog (Fig.\,\ref{fig:pulsar_examples}f). The coincidence of the pulsar with one of the two indicates that these sources are not related to each other, and that the Southern source is the LoTSS counterpart to PSR\,J0227+3356. 

\emph{J1336+3414:} In two separate LoTSS pointings, P204+32 and P204+35, there is a clear patch of emission at the location of PSR\,J1336+3414 (Fig.\ref{fig:pulsar_examples}e) though no source has been catalogued in LoTSS. A flux density of $\sim0.5$\,mJy is measured for the pulsar. This flux is a $\sim\,4\sigma$ detection, which together with the perturbing nearby source, explains the absence of the pulsar in the LoTSS catalog.

\emph{J2155+2813:} This pulsar is offset by $8\farcs4$ from the 4.7\,mJy LoTSS source ILT\,J215515.22+281309.3. While this is well outside the 99\% confidence error region of the pulsar based on the pulsar timing position by \citet{2004ApJ...600..905L}, the similarity between the LoTSS flux density and those obtained from earlier LOFAR time-domain observations (\citealt{LOTAAS} at $4.8\pm2.2$\,mJy and \citealt{Bilous+2016} at $9.6\pm4.8$\,mJy), the absence of another radio source at the pulsar position, and the absence of multi-wavelength counterparts in WISE/Panstarss at either location, leads us to consider ILT\,J215515.22+281309.3 as the counterpart to PSR\,J2155+2813. Future timing observations would have to confirm whether the timing ephemeris of \citet{2004ApJ...600..905L} is in error. We rule out proper motion, because the known distance to the pulsar (\citealt{2004ApJ...600..905L}, \citealt{Pulsar_catalog}) would imply an unrealistically high tangential velocity.

Two sources showed anomalies in the images:

\emph{B0823+26:} This pulsar is surrounded by clear imaging artifacts (Fig.\,\ref{fig:pulsar_examples}h) that are  not seen around other nearby sources in this LoTSS pointing. We hypothesise that these artifacts are caused by the intermittent behaviour (nulling) of the pulsar. The pulsar is known to turn on/off on a timescale of minutes and hours (\citealt{2012MNRAS.427..114Y}, \citealt{2015MNRAS.451.2493S}), which is on the same scale as the exposure time. We expect that due to the intermittent behaviour, the UV-coverage for this source is incomplete, which conflicts with the assumption of continuous emission of the \texttt{CLEAN} algorithm, leading to imaging artifacts (the surrounding bright spots) and the elliptical shape of the pulsar (Fig.\,\ref{fig:pulsar_examples}h). Analysis of the visibilities at higher time resolution could confirm this hypothesis. Furthermore, the flux density we measure is an order of magnitude lower than was measured at 150\,MHz in the literature (\citealt{Bilous+2016}, \citealt{Frail+2016}, \citealt{LOTAAS}). If the pulsar was not emitting as often during the LoTSS observation as it did during the literature observations, that would explain this discrepancy.

\emph{J1243+3946:} The counterpart to PSR\,J1243+3946 has an extension towards the East. Since all sources in the vicinity of the pulsar have this extension (Fig.\,\ref{fig:pulsar_examples}b), we attribute this to an issue with the direction dependent phase- calibration of this pointing.

Lastly, we note that the LoTSS positions of PSRs\,J0006+1834, B0153+39, J1059+6459 and J2355+2246 improve upon values available in the pulsar catalog, which have positional uncertainties of order $1\arcsec$.

\subsection{Non-detections}\label{ss:non-detections}
We did not find matches for 15 of 95 pulsars with accurate positions in v2.2.0 of the ATNF pulsar catalog. In this section, we discuss on an individual basis why we do not detect them. 

For two pulsars, no radio emission is expected, as PSR\,J1605+3249 is the X-ray dim isolated neutron star RX\,J1605.3+3249 \citep{2007Ap&SS.308..181H}, for which no radio emission has yet been detected (e.g.\ \citealt{2009ApJ...702..692K}), while PSR\,J1627+3219 is a millisecond pulsar associated with $\gamma$-ray source 4FGL\,J1627.7+3219 and detected through $\gamma$-ray pulsations and has no radio detection \citep{2023ApJ...958..191S}.

Six known radio pulsars in the globular clusters M3 (PSRs\,J1342+2822B and D) and M13 (PSRs\,J1641+3627C-F) are also not detected in LoTSS-DR2 imaging. As these pulsars have been discovered through deep targeted periodicity surveys, the sensitivity limits are significantly deeper than those of wide-field untargeted surveys. This is confirmed by the flux density measurements at 1400\,MHz for 5 of these pulsars from the pulsar catalog, which are in the range of 0.01 to 0.03\,mJy. Assuming a power-law spectra with $\alpha=-1.6$, this would yield $S_{144}\approx0.4$ to 1.1\,mJy and close to the LoTSS detection threshold.

There are 4 radio emitting pulsars (PSRs\,J0211+4235, J1048+5349 J1502+4653 and J2307+2225) which could be detected in LoTSS imaging.

\emph{J0211+4235:} The pulsar was discovered with FAST \citep{2023MNRAS.522.5152W} and a flux density of 109\,\textmu Jy was reported at 1.25\,GHz. Given the $5\sigma$ rms noise of 0.5\,mJy at 144\,MHz at the location of the pulsar, this constrains the spectral index of the source to $\alpha \geq -0.7$ assuming a power law spectrum.

\emph{J1502+4653:}
This pulsar was discovered with FAST \citep{2021MNRAS.508..300C} at 1.36\,GHz. With the given system specifications in \citet{2021MNRAS.508..300C}, the integration time for PSR\,J1502+4653 and the radiometer equation, we estimate that the $5\sigma$ sensitivity of the FAST observation was 23.5\,\textmu Jy. This constrains the spectral index to $\alpha \geq -1.4$ for a power law spectrum. 

\emph{J1048+5349:} 
This pulsar has been detected with the CHIME/Pulsar system \citep{2021ApJS..255....5C}, which has a minimum flux density limit of 0.6\,mJy in the frequency range of $400\sim800$\,MHz \citep{Dong+2023}. Given this minimum sensitivity, we would expect to detect this pulsar with LoTSS imaging for a large range of spectral indices. 

\emph{J2307+2225:} This pulsar has previously been detected in LOFAR time-domain observations \citep{Bilous+2016}, hence we expected to detect the pulsar in LoTSS imaging. At 149\,MHz its flux density is $0.9\,\pm\,0.7$\,mJy, which is above the detection limit of LoTSS. Upon visual inspection, we do not detect a low-significance, uncatalogued source in LoTSS images. Continued timing observations with the Lovell telescope (B.\ Stappers et al., private communication) confirm that the values in PSRCAT are accurate.

The final three known radio pulsars (PSRs\,J0139+3326, J1006+3015 and J2237+2828) that have not been detected in LoTSS-DR2 imaging are known Rotating Radio Transients (RRATs; \citealt{2006Natur.439..817M,2011BASI...39..333K}); these pulsars are detected by occasional individual pulses instead of periodic emission and hence will have low integrated flux densities that are likely below the LoTSS detection limit.

\emph{J0139+3326:} 
Detections of single pulses from this RRAT have been reported by \citet{2018ARep...62...63T}, were followed up in \citet{2021A&A...647A.191B} and later detected by \citet{LOTAAS} as well. According to \citet{LOTAAS}, the pulsar has an estimated flux density at 135\,MHz of 3.6\,mJy, which is above the detection threshold of LoTSS, yet we do not detect it. As reported by \citet{Michilli+2020}, the RRAT only sporadically emits single pulses, without a periodic signal that can be distinguished from noise. 

\emph{J1006+3015:}
The pulsar has been detected before by \citet{2022MNRAS.517.1126S}, and by \citet{McKenna+2024} using Irish LOFAR. The pulsar emits bright pulses on the order of 1\,hr$^{-1}$, which are likely to be averaged out by the long observation time of LoTSS. This is confirmed by ongoing, preliminary, analysis of the 8\,s LoTSS sub-integrations, which reveal 8 detections over the 8\,hr observation (S.\,Ranguin and R.\,Thomas, private communication).

\emph{J2237+2828:}
The pulsar was discovered with $\sim$ 70 hours of observation using CHIME \citep{Dong+2023} and was also detected by \citet{McKenna+2024}. It has been measured to emit bright pulses on the order of $1\sim2$ hours by \citet{Dong+2023} and \citet{2025MNRAS.537.1070T}, while no fainter pulses have been detected. 

\subsection{Flux densities}
The flux densities listed in Table\,\ref{tab:detections} expand upon the flux densities present in the literature. For 7 pulsars, image-domain flux densities at 150\,MHz are available from \citet{Frail+2016}. Additionally, \citet{TULLIP} presented flux density measurements for 6 pulsars, but since these based on pre-publication LoTSS observations, they are identical to those presented here. There are time-domain flux density measurements available in the literature at or around observing frequencies of 150\,MHz for 58 out of 80 pulsars detected with LoTSS. The literature fluxes are as follows: measurements from LOFAR observations as part of the normal pulsar and millisecond pulsar censuses by \citet{Bilous+2016} and \citet[][both at 150\,MHz]{Kondratiev+2016}, redetected pulsars as part of the LOTAAS survey \citep[][at 135\,MHz]{LOTAAS}, and measurements from timing observations of pulsars discovered in LOTAAS \citep[][at 150\,MHz, where we take the average of the 129 and 167\,MHz measurements]{2020MNRAS.492.5878T,Michilli+2020,2023A&A...669A.160V}.

\begin{figure}
    \centering
   \includegraphics[width=\columnwidth]{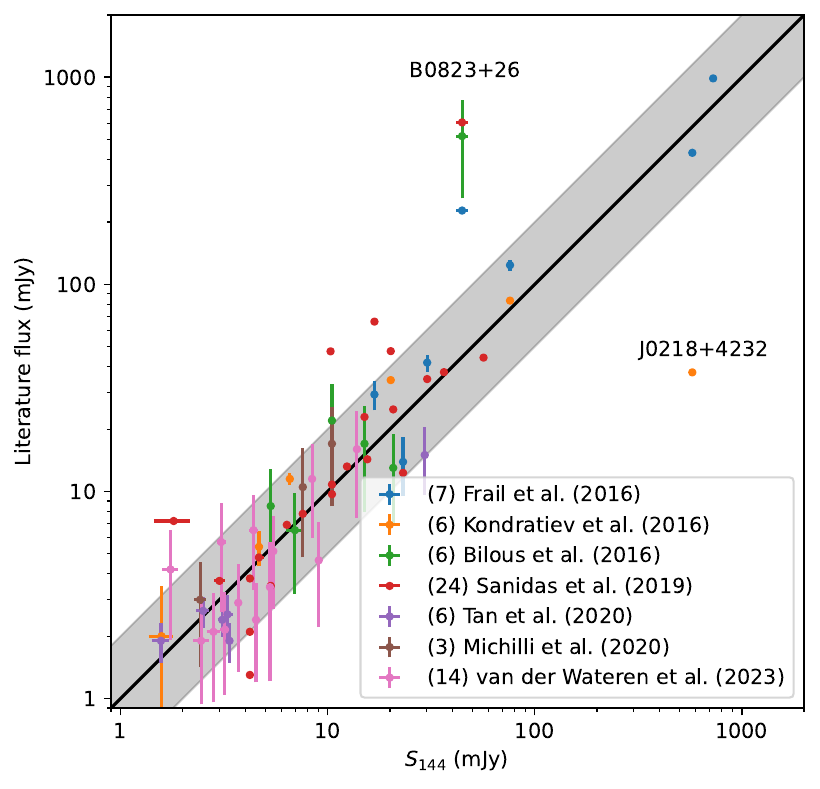}
   \caption{Comparison of LoTSS flux densities from Table\,\ref{tab:detections} at 144\,MHz (horizontal) to 150\,MHz flux density measurements from image-domain observations by \citet{Frail+2016} and time-domain observations by \citet{Bilous+2016,Kondratiev+2016,2020MNRAS.492.5878T,Michilli+2020,2023A&A...669A.160V} and \citet{LOTAAS} at 135\,MHz. The grey region represent fluxes ratios between factors 0.5 and 2.}
   \label{fig:flux_comparison}
\end{figure}

In general, the LoTSS flux densities are consistent with literature values for the majority of pulsars, see Fig.\,\ref{fig:flux_comparison}. Clear outliers are PSR\,B0823+26, which is a known nulling pulsar and for which measurements are available by \citet{Frail+2016,Bilous+2016} and \citet{LOTAAS}, and PSR\,J0218+4232, which has a matching flux density with \citet{Frail+2016} but a large difference compared to \citet{Kondratiev+2016}. For 5 of the LOTAAS redetections the time-domain flux densities are at least $2\times$ brighter or fainter than the LoTSS values. Since the LOTAAS flux densities are determined from search observations whose pointings were offset from the pulsar positions, \citet{LOTAAS} used a simple beam model to correct the observed flux densities. For 3 of the 5 pulsars the pointing offset was considerable (at least 2 beamwidths), which may explain the differences. For the 2 remaining LOTAAS redetections (PSRs\,J0944+4106 and J1059+6459) the pointing offsets are smaller, and we note that GBNCC observations at 350\,MHz by \citet{2018ApJ...859...93L} indicate these pulsars may display nulling.

\subsection{Pulsars in polarized emission}\label{ssec:polarization}
The LoTSS DR2 survey observations have been used to search for radio sources with polarized emission. \citet{O'Sullivan+2023} detected 2461 linearly polarized radio sources, of which 25 were matched to known radio pulsars, while \citet{V-LoTSS} detected 68 circularly polarized sources, with 24 being matched to known radio pulsars. A total of 12 radio pulsars are detected in both linear and circular polarization by \citet{O'Sullivan+2023} and \citet{V-LoTSS}. We compare these polarized detections to our sample of 80 known radio pulsars detected in the LoTSS survey through total intensity (Stokes I). We note that the pre-publication catalogues of \citet{O'Sullivan+2023} and \citet{V-LoTSS} were input to the work of \citet{TULLIP} that led to the discovery of new radio pulsars or independent re-discoveries of known but poorly localized radio pulsars using targeted pulsation searches of unidentified polarized point sources. These pulsars are included in this analysis as well as those of \citet{O'Sullivan+2023} and \citet{V-LoTSS}.

All 25 linearly polarized pulsars identified by \citet{O'Sullivan+2023} are also identified in LoTSS DR2 Stokes I images, though PSR\,B1112+50 has been mis-identified with the bright source offset and unrelated to the pulsar (further elaborated upon in Sect.\,\ref{subsec: Detections}). In the case of the 24 circularly polarized pulsars from \citet{V-LoTSS}, all but one are identified in LoTSS DR2, with the exception of PSR\,J0742+4110 which is not in the LoTSS DR2 Stokes I footprint. The known radio pulsars also identified from their polarized properties are marked in Table\,\ref{tab:detections}.

\begin{figure*}
    \centering
    \includegraphics[width=\textwidth]{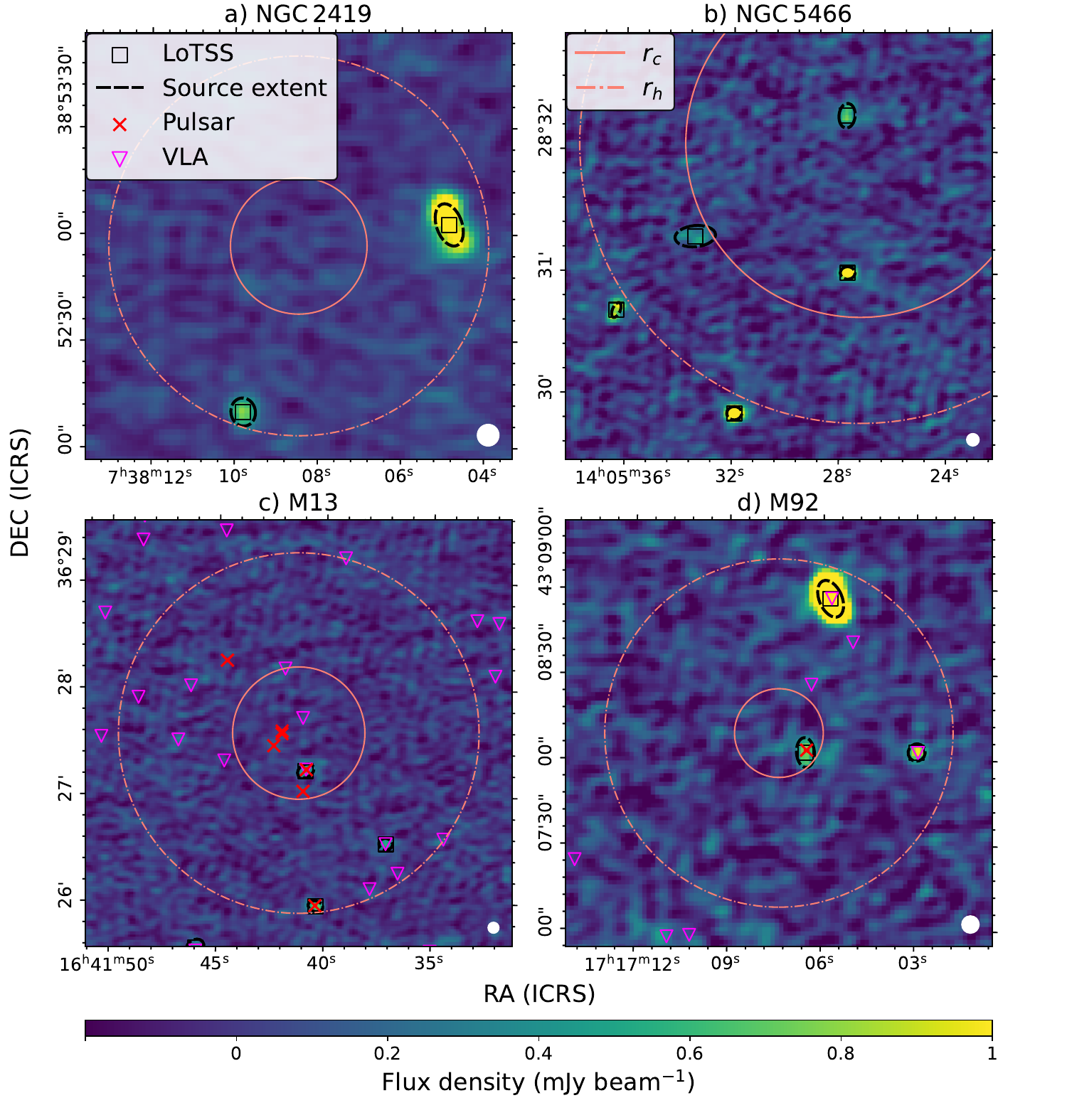}
    \caption[LoTSS images (Stokes I) of globular clusters]{LoTSS DR2 images in Stokes I for 7 globular clusters. The image in panel a) has a field-of-view of 2\arcmin\,$\times$\,2\arcmin. Panel b) has a field-of-view of 3\farcm5\,$\times$\,3\farcm5, and panels c) and d) have a field-of-view of 2.5\arcmin\,$\times$\,2.5\arcmin and 4\arcmin\,$\times$\,4\arcmin\, respectively.
    Positions of LoTSS sources are indicated with a black box, pulsar positions are denoted with the red cross and positions of VLA sources \citep{2020ApJ...903...73S} are shown with magenta triangles. The core- and half-light radii of globular clusters are plotted in salmon with solid and dash-dotted circles respectively. The FWHM of the synthesized LoTSS beam of 6\arcsec\,$\times$\,6\arcsec\, is shown in the bottom-right corners.
    } \label{fig:globular clusters}
\end{figure*}

As a result, of the 80 pulsars detected in total intensity, 35 stand out due to their polarization properties. As the detection of polarized emission depends both on the polarization fraction and the Stokes I flux density, detecting fainter sources through polarization is less likely. Pulsars have an average of 20\% and 10\% absolute linear and circular polarization fractions, respectively \citep{Lorimer+Kramer2012}. Hence, an average pulsar needs a brightness of $S_{144} \geq 2.5-5.0$\,mJy to be detected in the respective polarizations, which matches what we observe in our sample (2.0\,mJy for linear and 4.9\,mJy for circular). If we conservatively limit our sample to Stokes I flux densities of $S_{144}\ga 4$\,mJy, we find that 34 out of 51 pulsars (67\%) are detectable through their polarization properties, with approximately equal fractions of either linear or either circular or both linear and circular polarization. We note that the bright millisecond pulsar PSR\,J0218+4232 ($S_{144} = 577$\,mJy) was not detected as a polarized source by \citet{O'Sullivan+2023} or \citet{V-LoTSS}, while it has previously been reported to have a linear polarization fraction of 40\% \citep{1995ApJ...455L..55N} up to 48\% \citep{TULLIP}, which far exceeds the detection limit of \citet{O'Sullivan+2023}. The high brightness of PSR\,J0218+4232 likely caused artificial polarized sources, which prevent its detection. Similarly, PSR\,J0815+4611 is not detected by \citet{O'Sullivan+2023}. The pulsar has a linear polarization fraction of \~50\% \citep{2015A&A...583A.137J} which translates to a flux density of twice the detection limit. Comparable to PSR\,J0218+4232, nearby bright sources could be causing artifacts that perturb the detection of PSR\,J0815+4611.

\subsection{Globular Clusters}
Radio (millisecond) pulsars are abundantly present in the dense stellar environments of globular clusters due to the combined effect of high stellar encounter rates and mass segregation where neutron stars tend to move to the cluster center. As a result, neutron stars can be repeatedly spun up by mass transfer from one or more binary companions \citep[e.g.][]{2014A&A...561A..11V}.

The 2010 version of the \citet{harris1996} globular cluster catalog shows that the LoTSS DR2 footprint includes 7 Galactic globular clusters; M3, M13, M92, Ko\,2, Pal\,4, NGC\,2419 and NGC\,5466. Radio pulsars are known in M3, M13 and M92, with 6 in M3, of which 2 have accurate positions, but none are detected in LoTSS DR2. M13 has 6 pulsars, all with accurate positions of which 2 are detected; PSRs\,B1639+36A and B1639+36B. PSR\,J1717+4308A in M92 is the only known pulsar in that cluster, and has an accurate position and is detected in LoTSS DR2. See Sect.\,\ref{ss:non-detections} for the non-detection of the pulsars in M3 and M13.

No radio pulsars are known in Ko\,2, Pal\,4, NGC\,2419 and NGC\,5466\footnote{\url{https://www3.mpifr-bonn.mpg.de/staff/pfreire/GCpsr.html}}. As these 4 globular clusters have lower central luminosity densities ($\rho_0=10^{0.1-1.6}$\,L$_\odot$\,pc$^{-3}$; \citealt{harris1996}) compared to M3, M13 and M92 ($\rho_0=10^{3.5-4.3}$\,L$_\odot$\,pc$^{-3}$), the lower stellar density and hence lower stellar encounter rate would predict that fewer neutron stars would be created due to dynamical interactions in these clusters \citep{2014A&A...561A..11V}. Additionally, Ko\,2, Pal\,4, NGC\,2419 and NGC\,5466 are more distant ($d=16$ to 83\,kpc) than M3, M13 and M92 ($d=7$ to 10\,kpc), and hence any radio pulsars in those clusters would be fainter.

Unidentified LoTSS point sources in or near globular clusters may correspond to unlocalized or even undiscovered radio pulsars. We do not detect any sources within the half-light radii of globular clusters Ko\,2, Pal\,4 and M3. Figure\,\ref{fig:globular clusters} shows the LoTSS images and catalogued sources in NGC\,2419, NGC\,5466, M13 and M92, for which sources are detected within the half-light radii from \citet{harris1996}. These clusters are also covered in VLITE commensal continuum imaging observations by \citet{2024ApJ...969...30M} at 340\,MHz (NGC\,2419 and NGC\,5466), deep targeted VLA continuum observations by \citet{2020ApJ...903...73S} at 5.0 and 7.2\,GHz (M3, M13 and M92) and X-ray observations with Chandra/ACIS \citep{MAVERIC} (M13, M92). We do not consider ILT\,J073804.85+385302.8 in NGC\,2419, ILT\,J140527.79+283217.3 and J140533.42+283117.3 in NGC\,5466 and ILT\,J171705.77+430856.7 in M92 as these are resolved sources in LoTSS. We note that the latter corresponds to M92-VLA1 from \citet{2020ApJ...903...73S}. 

The unresolved LoTSS source ILT\,J073809.79+385209.9 in NGC\,2419 has no counterparts at other radio frequencies or in X-rays, and the \citet{2024ApJ...969...30M} non-detection is not constraining the spectral index. No pulsars have been reported by \citet{2025ApJS..279...51L} using targeted time-domain periodicity search with FAST ($S_{1400}>1.59$\,\textmu Jy). ILT\,J140527.71+283059.9 in NGC\,5466 is unresolved and detected at 340\,MHz as VLITE-A\,J140527.6+283100 by \citet{2024ApJ...969...30M}, who have used the LoTSS DR2 flux density to show it has a flat spectrum ($\alpha=-0.1\pm0.5$). The unresolved source ILT\,J164137.13+362633.6 in M13 matches M13-VLA4 and the $S_{144}=0.77\pm0.19$\,mJy LoTSS flux density is consistent with the $\alpha=-0.94\pm0.31$ spectral index determined from the 5.0 and 7.2\,GHz VLA detections by \citet{2020ApJ...903...73S}. With $\alpha=-2.74\pm0.44$, M92-VLA12 has one of the steepest spectral indices from the \citet{2020ApJ...903...73S} sample and would have $S_{144}>90$\,mJy in LoTSS. Instead, it is detected at $S_{144}=1.0\pm0.2$\,mJy as ILT\,J171702.96+430803.0, suggesting the spectrum is significantly flatter ($\alpha\sim-1.1$) or follows a broken power-law. Hence, none of these unresolved sources have (known) spectral indices that would suggest they would be promising targets for targeted pulsar periodicity searches. The lack of X-ray counterparts to these radio sources supports this.

The absence of LoTSS counterparts to the radio sources reported by \citet{2020ApJ...903...73S} at 5.0 and 7.2\,GHz located within the half-light radii of M3, M13 and M92 indicates that if these sources would have power-law spectra, these would have to be relatively shallow, with $\alpha\gtrsim -1.5$, and mostly inconsistent with those of radio pulsars whose spectra follow a distribution with $\alpha=-1.6\pm1.0$ \citep{2018MNRAS.473.4436J}. Continuum observations at intermediate frequencies would be required to rule out broken power-law or curved spectral models.

\section{Discussion and conclusions}\label{sec:discussion_and_conclusions}
\subsection{Recovery fraction}
\begin{figure}
    \centering
   \includegraphics[width=\columnwidth]{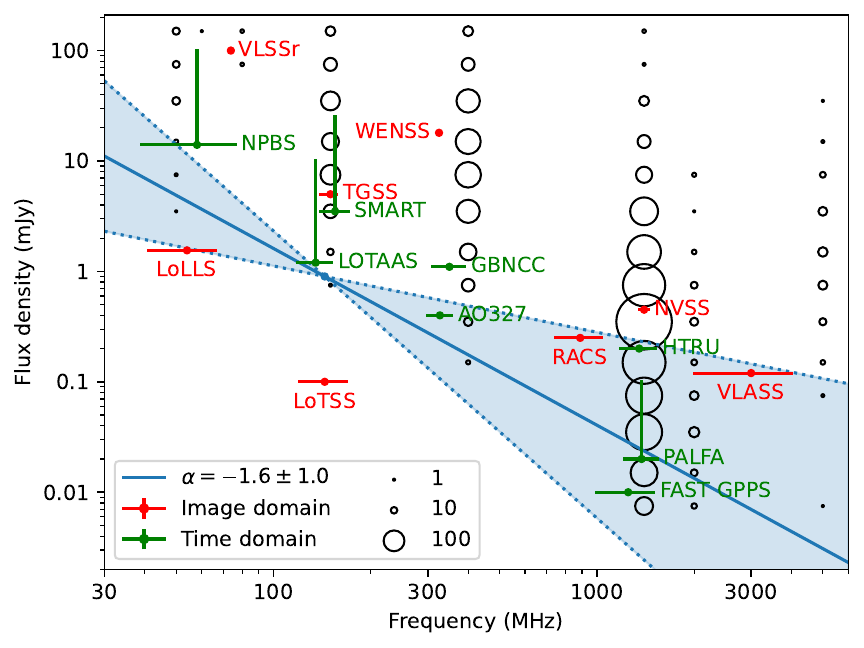}
   \caption{A comparison of survey sensitivity and frequency coverage
     of various time domain pulsar surveys (green) and wide area
     continuum image domain surveys (red) against distributions of
     pulsar flux density measurements from the pulsar catalog (open
     circles). An example power law spectrum $S_\nu\propto\nu^\alpha$
     with $\alpha=-1.6\pm1.0$ for $S_{144}=0.9$\,mJy is shown for
     reference. (References: NPBS; \citealt{2025A&A...693A..96B}, LOTAAS;
     \citealt{LOTAAS}, SMART; \citealt{2023PASA...40...20B}, AO327;
     \citealt{2013ApJ...775...51D}, GBNCC;
     \citealt{2014ApJ...791...67S}, FAST GPPS;
     \citealt{2021RAA....21..107H}, PALFA;
     \citealt{2006ApJ...637..446C}, HTRU;
     \citealt{2010MNRAS.409..619K}, LoLLS;
     \citealt{2021A&A...648A.104D}, VLSSr; 
     \citealt{2014MNRAS.440..327L}, LoTSS; 
     \citealt{Shimwell+17},
     TGSS; \citealt{2017A&A...598A..78I}, WENSS;
     \citealt{1997A&AS..124..259R}, RACS;
     \citealt{2020PASA...37...48M}, NVSS;
     \citealt{1998AJ....115.1693C}, VLASS; \citealt{Lacy+2020}).}
   \label{fig:fluxdensities}
\end{figure}

Figure\,\ref{fig:fluxdensities} compares the sensitivity limits and frequency coverage of several time domain pulsar surveys and wide field image domain surveys against observed flux densities of the radio pulsar population from the pulsar catalog. At observing frequencies around 150\,MHz, survey sensitivity of the LoTSS image-domain survey should allow for the detection of all known radio pulsars with flux density measurements at that frequency. While the sensitivity limits of time-domain pulsar surveys at L-band frequencies with the large Arecibo (PALFA; \citealt{2006ApJ...637..446C}) and FAST telescopes (GPPS; \citealt{2021RAA....21..107H}) are almost an order of magnitude lower than that of LoTSS, the typically steep radio spectra of pulsars (e.g.\ $S_\nu\propto\nu^\alpha$ with $\alpha=-1.6\pm1.0$; \citealt{2018MNRAS.473.4436J}) would result in 150\,MHz flux densities above the LoTSS sensitivity limit for the majority of the observed pulsar flux density distributions at higher frequencies and over a significant range of power-law spectral indices. As such, we would expect that the majority of radio pulsars would be detectable in LoTSS continuum imaging.

This is confirmed by our catalog matching, as we find that of the 95 pulsars located within the DR2 footprint for which their catalogued positional information allows unambiguous matching, 80 pulsars are recovered in LoTSS. Of the 15 pulsars that were not recovered, there are valid reasons for their non-detection, either due to the pulsars being radio quiet (2 cases), known to be faint at higher frequencies (8 cases) or being RRATs that have low integrated flux densities due to their occasional bright pulses (3 cases). Only in two cases (PSRs\,J1048+5349 and J2307+2225) were non-detections unexpected.

As a result, we recover over 86\% (up to 97\% when accounting for pulsars which are known RRATs, radio-quiet and/or faint) of the radio pulsars with accurate locations (better than $20\arcsec$) that coincide with the LoTSS DR2 footprint. At 150\,MHz, this fraction can be compared to the 23\% recovery fraction (288 out of 1238) of \citet{Frail+2016} using the TGSS ADR1 image-domain survey \citep{2017A&A...598A..78I} and 8\% (83 out of 1000) from \citep{Sett+2024} with the MWA telescope (SMART survey; \citealt{2023PASA...40...20B}) due to their higher (3-8\,mJy) sensitivity limits. At higher frequencies, 661 out of 2235 known radio pulsars (30\%) were recovered in the RACS survey with ASKAP at 888\,MHz \citep{2023ApJ...956...28A}, while the NVSS survey at 1.4\,GHz recovered 79 of the 516 radio pulsars (15\%) known at the time \citep{Kaplan1998}.

Conservatively, 51 of the detected pulsars are bright enough ($S_{144} \geq 4$\,mJy) to surpass the sensitivity of LoTSS in polarized emission (\citealt{O'Sullivan+2023}, \citealt{V-LoTSS}). We find that the resulting fraction of pulsars that are detectable in polarized emission of 68\% is high compared to the 48\% obtained by \citet{Frail+2024} in the MeerKAT survey of the Galactic bulge at L-band, where we note that \citet{Frail+2024} includes a larger number of false positives in their matches. Polarization leakage is similar for all polarization types (\citealt{Shimwell+22}, \citealt{2024MNRAS.528.2511T}), and polarization fractions for which a source is considered polarized are similar between \citet{Frail+2024}, \citet{O'Sullivan+2023} and \citet{V-LoTSS}.

\subsection{Flux densities}
Following the definition and recommendation from \citep{Shimwell+22} separating resolved from unresolved sources based on the ratio between integrated flux density and peak brightness, we find that the properties of the pulsars catalogued in LoTSS DR2 are all consistent with being unresolved. Since none of the matched pulsars are known to be associated with supernova remnants or pulsar wind nebulae, this matches expectations. Conversely, the measured flux densities represent the phase averaged pulsar emission and will not be skewed by radio emission from supernova remnants or pulsar wind nebulae \citep[e.g.][]{Frail+2016}.

Due to the long integration times (8\,h) and observing bandwidth (48\,MHz) used by LoTSS, the LoTSS observations are robust against scintillation, as scintillation generally occurs on timescales and bandwidths on the order of minutes in duration and several kHz to a MHz in bandwidth \citep[e.g][]{2022A&A...663A.116W}. As a result, the LoTSS flux densities generally match literature values, with the exception of PSRs\,B0823+26 and J0218+4232. 

The former is a known pulsar which displays long nulls -- the temporary absence of pulsations -- on timescales of minutes to hours \citep{2012MNRAS.427..114Y,2015MNRAS.451.2493S}. As a result, flux density measurements are impacted by the behaviour of the pulsar during observations, leading to large variations in the integrated flux density between observations. PSR\,J0218+4232 is a millisecond pulsar ($P=2.3$\,ms, $\mathrm{DM}=61.25$\,pc\,cm$^{-3}$) which is not known to be nulling, and the image-domain flux densities of $S_{144}=577(87)$\,mJy and $S_{150}=432(43)$\,mJy from LoTSS and TGSS ADR1 \citep{Frail+2016,2017A&A...598A..78I}, while consistent with each other, are significantly brighter than the LOFAR time-domain measurement of $S_{150}=38(19)$\,mJy \citep{Kondratiev+2016}. While \citet{Frail+2016} attributes the difference to scintillation, \citet{Kondratiev+2016} argues that the wide phase resolved pulse profile of PSR\,J0218+4232 could be affected by scattering, resulting in the pulsed flux density being underestimated in comparison to the integrated flux density. Since a significant fraction of the radio emission of PSR\,J0218+4232 is known to be unpulsed \citep{1995ApJ...455L..55N,2015MNRAS.453..828K} due to the pulsar being an aligned rotator \citep{1999ApJS..123..627S}, we consider it likely that this -- at least partially -- explains the large difference between the pulsed and total integrated flux densities.

The loss of pulsed flux due to scatter broadening in time-domain observations is expected at LOFAR frequencies. The sample of 73 newly discovered pulsars and 311 redetected known pulsars in the LOTAAS survey at 135\,MHz all have dispersion measures of $\mathrm{DM}\lesssim220$\,pc\,cm$^{-3}$, which is in agreement with sensitivity limits that account for temporal broadening of pulse profiles due to scattering (which approximately scales with dispersion measure) \citep{LOTAAS}. Since LoTSS DR2 covers high Galactic latitudes, the dispersion measures in our sample are all within that limit, with PSR\,J2155+2813 having the highest at $\mathrm{DM}=77.13$\,pc\,cm$^{-3}$, and hence for slow pulsars (e.g.\ $P>0.1$\,s), we do not expect differences between the observed time-domain and image-domain flux densities. For faster spinning pulsars, the effects of scatter broadening can still be significant, as may be the case for PSR\,J0218+4232. The time-domain flux densities of the other 5 millisecond pulsars (PSRs\,J1012+5307, J1544+4937, J1816+4510, J2214+3000, J2322+2057) investigated by \citet{Kondratiev+2016} are all consistent with the LoTSS image-domain flux densities, and \citet{Kondratiev+2016} does not consider their profiles to be scattering dominated. These 5 millisecond pulsars all have longer spin periods, and/or smaller dispersion measurements, compared to PSR\,J0218+4232 so the effects of scatter broadening would be expected to be less.

\subsection{Selection criteria for identifying pulsar candidates}
The high sensitivity and spatial resolution of the LoTSS survey offers the possibility of identifying promising pulsar candidates from the continuum sources detected as part of this survey. Given the characteristic properties of pulsars in the image domain as generally unresolved, steep radio spectrum, and linearly and/or circularly polarized sources that lack counterparts at other wavelengths, we use the properties of the sample of known pulsars in LoTSS DR2 to evaluate which properties provide the best means of identifying pulsar candidates.

Since the majority (92\%) of the radio sources in LoTSS DR2 are classified as unresolved \citep{Shimwell+22}, a selection based purely using unresolved source properties is not constraining.

The LoTSS survey offers flux density measurements at the center of the observed bandwidth, and hence determining in-band spectra for each radio source is not possible from the current data. As evident from Fig.\,\ref{fig:fluxdensities}, cross-matching LoTSS source positions to radio sources in other image-domain surveys is likely leading to detections for LoTSS sources that are bright and/or have a flat spectrum, with only the LoLLS, RACS and VLASS surveys offering the possibility of detecting counterparts to a large range in flux densities at 144\,MHz and spectral indices. Of the 80 radio pulsars recovered in LoTSS, only a few have counterparts in the surveys with the VLA that fully cover the LoTSS DR2 footprint, with 2 counterparts at 74\,MHz (VLSSr), 4 at 1400\,MHz (NVSS), and 3 at 3\,GHz (VLASS). Partial coverage is offered by RACS (up to $\delta<41$\degr), which has 1 counterpart and WENSS with 2 counterparts. The LOFAR LBA Sky Survey (LoLSS) at 54\,MHz is most promising due to its depth of $S_{54}\sim1$\,mJy\,beam$^{-1}$ \citep{LoLSS2021} and of the 8 known pulsars with accurate positions present in the 650\,deg$^2$ area released in LoLLS DR1 \citep{LoLSS2023}, 2 are detected (PSRs\,B1508+55 and J1552+5437), resulting in a 25\% recovery fraction. A cross matched catalog between LoLLS at 54\,MHz and LoTSS at 144\,MHz by \citet{2023A&A...674A.189B} reveals that the majority (99.5\%) of (non-artifact) radio sources in LoLSS have counterparts in LoTSS, with about 1.3\% having steep power-law spectral indices ($\alpha<-2$). A selection based on spectral properties between a cross matched catalog of LoLSS and LoTSS radio sources removes the majority of other sources (e.g. radio loud AGN that have typical spectral indices of $\alpha=-0.7$, e.g. \citealt{2017MNRAS.469.3468C}, \citealt{2019A&A...622A..17S}) and would yield hundreds to thousands of candidates. Candidates can be further constrained by removing sources that have been matched to optical counterparts by \citet{2023A&A...678A.151H}.

As discussed in Sect.\,\ref{ssec:polarization}, the surveys for linearly and circularly polarized emission from LoTSS DR2 sources by \citet{O'Sullivan+2023} and \citet{V-LoTSS} identified 35 of the 80 known radio pulsars by their polarization properties, making candidate selection based on polarization very promising. \citet{TULLIP} used this approach on pre-publication data from \citet{O'Sullivan+2023} and \citet{V-LoTSS} and selected pulsar candidates from linearly polarized sources with $S_{144}>2$\,mJy and $>7$\% polarization fraction and circularly polarized sources with $S_{144}>5$\,mJy and $>1$\% polarization fraction that were unresolved in Stokes\,I and did not have multi-wavelength counterparts that would identify them as likely galaxies. This yielded 9 linearly polarized candidates of which 3 were identified as radio pulsars, and 21 circularly polarized candidates, 5 of which were identified as radio pulsars.

Based on these results, selecting pulsar candidates based on polarization properties is the most promising approach to find new pulsars from wide-area image-domain surveys, though we expect that cross-matching LOFAR radio sources from LoLSS at 54\,MHz with LoTSS at 144\,MHz to select based on spectral properties will result in additional pulsar candidates, and provide some bias against pulsars that do not appear polarized. These approach matches that suggested by \citet{Sett+2023} based on studies with the MWA.

\subsection{Outlook}
The upcoming LoTSS DR3 release (Shimwell et al., in prep.) will cover the full Northern hemisphere above declination $\delta>0$\degr at 120--168\,MHz and will include the Galactic plane to offer coverage of another 1330 known radio pulsars, of which some 870 have accurate positional information in PSRCAT (v2.2.0). The recovery rate of known pulsars in the Galactic plane likely will be lower compared to LoTSS DR2, due to deep time-domain Galactic plane surveys at with the large Arecibo and FAST telescopes \citep{2006ApJ...637..446C,2021RAA....21..107H}, as their L-band sensitivities are high enough that only relatively steep pulsar spectra ($\alpha\lesssim-1.7$) are detectable with LoTSS at 144\,MHz. Additionally, the presence of resolved and diffuse structures such as supernova remnants, \ion{H}{II} regions and pulsar wind nebulae at low Galactic latitudes may impact the calibration of LoTSS visibilities, leading to reduced image fidelity.

Since pulsars at lower Galactic latitudes can have dispersion measures in excess of $\mathrm{DM}\gtrsim220$\,pc\,cm$^{-3}$, where scatter broadening at radio frequencies near 150\,MHz is observed to completely smear out pulsations of even non-recycled pulsars ($P\sim1$\,s; \citealt{LOTAAS}) in time-domain observations, the LoTSS image-domain survey would provide crucial flux density measurements to assess pulsar spectra of a large sample of radio pulsars at 144\,MHz.

Additionally, the LOFAR LBA Sky Survey (LoLSS) will cover the Northern hemisphere above declination $\delta>24$\degr in the 41--66\,MHz frequency range \citep{2021A&A...648A.104D}, and once released will provide additional flux density measurements and, when combined with LoTSS detections, image-based spectral measurements. Together, the complete LoTSS and LoLSS surveys will provide a unique dataset to investigate the spectral properties of radio pulsars as well as select pulsar candidates for time-domain follow up observations.

\begin{acknowledgements}
JRC acknowledges funding from the European Union via the European Research Council (ERC) grant Epaphus (project number 101166008). This paper made extensive use of the Python scientific software stack \citep{PER-GRA:2007}, and we acknowledge the developers of \texttt{numpy} \citep{numpy}, \texttt{matplotlib} \citep{matplotlib}, \texttt{psrqpy} \citep{psrqpy}, \texttt{scipy} \citep{scipy}, \texttt{astropy} \citep{astropy2013,astropy2018,astropy2022}, \texttt{aplpy} (\citealt{aplpy}, \citealt{aplpy2019}), \texttt{MOC} \citep{2014ivoa.spec.0602F} and \texttt{astroquery} \citep{2019AJ....157...98G}. 
\end{acknowledgements}

\bibliographystyle{aa}

\onecolumn

\begin{appendix}
\section{Pulsars detected in LoTSS DR2}

\begin{longtable}[c]{llrr|lrrcl}
\caption{Properties of pulsars and their matching source from LoTSS.} \label{tab:detections} \\
\hline \hline
Pulsar & Pol. &
       \multicolumn{1}{c}{$\mathrm{P}$} &
       \multicolumn{1}{c}{$\mathrm{DM}$} &
       ILT &
       \multicolumn{1}{c}{$\mathrm{\Delta \alpha}$} &
       \multicolumn{1}{c}{$\mathrm{\Delta \delta}$} &
       \multicolumn{1}{c}{$\mathrm{S_{144}}$} &
       \multicolumn{1}{l}{Field} \\
       & & 
       \multicolumn{1}{c}{(ms)} &
       \multicolumn{1}{c}{(pc/cc)}
       &
       &
       \multicolumn{1}{c}{($\arcsec$)} &
       \multicolumn{1}{c}{($\arcsec$)} &
       \multicolumn{1}{c}{(mJy)} & \\
\hline
\endfirsthead
\caption[]{continued.}\\
\hline \hline
Pulsar & Pol. &
       \multicolumn{1}{c}{$\mathrm{P}$} &
       \multicolumn{1}{c}{$\mathrm{DM}$} &
       ILT &
       \multicolumn{1}{c}{$\mathrm{\Delta \alpha}$} &
       \multicolumn{1}{c}{$\mathrm{\Delta \delta}$} &
       \multicolumn{1}{c}{$\mathrm{S_{144}}$} &
       \multicolumn{1}{l}{Field} \\
       & & 
       \multicolumn{1}{c}{(ms)} &
       \multicolumn{1}{c}{(pc/cc)}
       &
       &
       \multicolumn{1}{c}{($\arcsec$)} &
       \multicolumn{1}{c}{($\arcsec$)} &
       \multicolumn{1}{c}{(mJy)} & \\
\hline
\endhead
\hline
\endfoot

J0006$+$1834	&		&	694	&	11	&	J000604.63$+$183455.7	&	$+2\farcs43(3\farcs03)$	&	$+3\farcs28(4\farcs02)$	&	7.0(1.2)	&	P002+18	\\
J0039$+$3546	&		&	537	&	53	&	J003908.81$+$354616.7	&	$+0\farcs04(0\farcs25)$	&	$-0\farcs35(0\farcs26)$	&	5.5(0.8)	&	P009+36	\\
B0045$+$33	&	V	&	1217	&	40	&	J004833.97$+$341207.4	&	$+0\farcs13(0\farcs51)$	&	$+0\farcs61(0\farcs34)$	&	20.7(3.1)	&	P013+34	\\
J0146$+$3055	&		&	938	&	25	&	J014640.97$+$305520.3	&	$+0\farcs72(0\farcs60)$	&	$-1\farcs23(0\farcs52)$	&	3.8(0.6)	&	P027+31	\\
J0154$+$1833	&	L	&	2.365	&	20	&	J015436.89$+$183350.8	&	$-0\farcs04(0\farcs32)$	&	$-0\farcs19(0\farcs28)$	&	4.2(0.7)	&	P029+19	\\
B0153$+$39	&		&	1812	&	59.83	&	J015654.94$+$394929.8	&	$+3\farcs50(4\farcs51)$	&	$-1\farcs35(3\farcs25)$	&	5.3(0.8)	&	P028+41	\\
J0158$+$2106	&	L	&	505	&	20	&	J015845.99$+$210646.7	&	$+0\farcs10(1\farcs53)$	&	$+0\farcs29(0\farcs31)$	&	4.3(0.7)	&	P029+21	\\
J0218$+$4232	&	L\tablefootmark{a}	&	2.323	&	61.25	&	J021806.34$+$423217.4	&	$+0\farcs25(0\farcs20)$	&	$-0\farcs11(0\farcs20)$	&	577.3(86.7)	&	P035+41	\\
J0220$+$3626	&	L/V	&	1030	&	40	&	J022042.12$+$362655.7	&	$-0\farcs02(0\farcs37)$	&	$-0\farcs07(0\farcs29)$	&	24.5(3.7)	&	P035+36	\\
J0227$+$3356	&		&	1240	&	27	&	--\tablefootmark{b}	&	--	&	--	&	2.1(0.4)	&	P035+34	\\
B0751$+$32	&	L/V	&	1442	&	40	&	J075440.65$+$323156.5	&	$+0\farcs40(0\farcs27)$	&	$-0\farcs17(0\farcs31)$	&	15.1(2.3)	&	P118+32	\\
J0811$+$3729	&		&	1248	&	15	&	J081115.14$+$372915.0	&	$-0\farcs52(0\farcs39)$	&	$-1\farcs27(0\farcs68)$	&	1.7(0.3)	&	P122+37	\\
J0815$+$4611	&	L\tablefootmark{c}	&	434.2	&	11	&	J081559.49$+$461153.4	&	$-0\farcs33(0\farcs26)$	&	$-0\farcs19(0\farcs26)$	&	10.5(1.6)	&	P122+47	\\
B0823$+$26	&	V	&	530.7	&	19	&	J082651.49$+$263720.4	&	$+0\farcs47(0\farcs38)$	&	$+0\farcs31(0\farcs43)$	&	44.6(7.3)	&	P126+27	\\
J0828$+$5304	&		&	14	&	23	&	J082825.68$+$530443.2	&	$+0\farcs17(0\farcs25)$	&	$-0\farcs14(0\farcs26)$	&	4.5(0.7)	&	P126+52	\\
J0854$+$5449	&		&	1233	&	18.84	&	J085425.70$+$544928.8	&	$+0\farcs21(0\farcs30)$	&	$-0\farcs04(0\farcs32)$	&	1.9(0.3)	&	P133+55	\\
J0857$+$3349	&		&	243	&	24	&	J085707.09$+$334918.3	&	$-0\farcs26(0\farcs43)$	&	$-1\farcs35(0\farcs68)$	&	2.4(0.4)	&	P135+34	\\
B0917$+$63	&	L/V	&	1568	&	13.15	&	J092114.15$+$625413.9	&	$-0\farcs13(0\farcs24)$	&	$-0\farcs01(0\farcs28)$	&	10.5(1.6)	&	P141+62	\\
J0925$+$6103	&		&	5.983	&	21.64	&	J092517.58$+$610304.2	&	$-0\farcs27(0\farcs39)$	&	$-0\farcs15(0\farcs37)$	&	1.5(0.3)	&	P141+62	\\
J0928$+$3039	&		&	2092	&	22	&	J092859.38$+$303926.9	&	$-0\farcs20(0\farcs32)$	&	$-0\farcs44(0\farcs39)$	&	3.1(0.5)	&	P141+29	\\
J0935$+$3312	&		&	962	&	18	&	J093507.82$+$331237.1	&	$-0\farcs32(0\farcs27)$	&	$-0\farcs53(0\farcs29)$	&	3.7(0.6)	&	P142+32	\\
J0944$+$4106	&	L	&	2229	&	21	&	J094418.14$+$410604.1	&	$+0\farcs02(0\farcs26)$	&	$+0\farcs37(0\farcs28)$	&	4.2(0.6)	&	P146+42	\\
J1012$+$5307	&	V	&	5.256	&	9.02	&	J101233.44$+$530702.2	&	$+0\farcs00(0\farcs21)$	&	$-0\farcs10(0\farcs21)$	&	20.2(3.0)	&	P151+52	\\
J1017$+$3011	&		&	452.8	&	27	&	J101736.30$+$301146.0	&	$-0\farcs11(0\farcs56)$	&	$+0\farcs07(0\farcs52)$	&	2.5(0.4)	&	P153+30	\\
J1049$+$5822	&	L	&	727.6	&	12.36	&	J104937.86$+$582217.6	&	$+0\farcs34(1\farcs23)$	&	$-0\farcs54(0\farcs94)$	&	2.0(0.3)	&	P161+60	\\
J1059$+$6459	&		&	3631	&	19.3	&	J105927.36$+$645932.0	&	$+0\farcs92(0\farcs40)$	&	$-0\farcs15(2\farcs67)$	&	4.2(0.6)	&	P162+65	\\
B1112$+$50	&		&	1656	&	9.19	&	J111537.51$+$503006.1\tablefootmark{d}	&	--	&	--	&	21(4)	&	P10Hetdex	\\
J1239$+$3239	&	L	&	4.701	&	16.86	&	J123927.33$+$323923.4	&	$-0\farcs20(0\farcs21)$	&	$-0\farcs09(0\farcs21)$	&	20.6(3.1)	&	P188+32	\\
J1243$+$3946	&		&	1310	&	29	&	J124303.34$+$394609.7	&	$+1\farcs91(0\farcs46)$	&	$-0\farcs53(0\farcs34)$	&	5.7(0.9)	&	P191+40	\\
J1303$+$3815	&		&	396	&	19	&	J130319.35$+$381503.2	&	$-0\farcs06(0\farcs29)$	&	$-0\farcs10(0\farcs29)$	&	3.2(0.5)	&	P197+37	\\
J1327$+$3423	&		&	41.51	&	4	&	J132707.56$+$342337.9	&	$-0\farcs17(0\farcs21)$	&	$-0\farcs26(0\farcs21)$	&	114.3(17.2)	&	P201+35	\\
J1336$+$3414    & &    3013   &   8.5 &   --\tablefootmark{e}  &   --  &   --  &   0.5(0.3)   &   P204+35 \\
J1343$+$6634	&	L	&	1394	&	30.03	&	J134359.33$+$663425.3	&	$-0\farcs44(0\farcs38)$	&	$-0\farcs31(0\farcs24)$	&	7.6(1.1)	&	P205+67	\\
J1427$+$5211	&		&	996	&	25.37	&	J142707.74$+$521111.5	&	$-0\farcs23(0\farcs37)$	&	$+0\farcs43(0\farcs36)$	&	2.8(0.5)	&	P214+52	\\
B1508$+$55	&	L	&	739.7	&	19.62	&	J150925.49$+$553131.7	&	$+0\farcs61(0\farcs21)$	&	$-0\farcs17(0\farcs21)$	&	727.8(109.3)	&	P227+55	\\
J1518$+$4904	&	L/V	&	40.93	&	11.61	&	J151816.80$+$490434.1	&	$-0\farcs05(0\farcs22)$	&	$-0\farcs05(0\farcs22)$	&	10.3(1.6)	&	P231+50	\\
J1529$+$4050	&		&	476	&	7	&	J152916.53$+$405057.8	&	$-0\farcs03(0\farcs25)$	&	$-0\farcs08(0\farcs24)$	&	8.4(1.3)	&	P231+40	\\
J1541$+$4703	&		&	277.7	&	19.4	&	J154105.54$+$470304.0	&	$-0\farcs04(0\farcs89)$	&	$-0\farcs39(0\farcs53)$	&	0.7(0.2)	&	P236+48	\\
J1544$+$4937	&	L/V	&	2.159	&	23	&	J154404.49$+$493755.3	&	$-0\farcs02(0\farcs24)$	&	$-0\farcs01(0\farcs23)$	&	6.6(1.0)	&	P235+50	\\
J1552$+$5437	&	V	&	2.428	&	22.9	&	J155253.34$+$543705.6	&	$-0\farcs04(0\farcs25)$	&	$+0\farcs14(0\farcs25)$	&	4.9(0.7)	&	P236+55	\\
J1602$+$3901	&	V	&	4	&	17.26	&	J160218.82$+$390101.5	&	$+0\farcs21(1\farcs22)$	&	$+0\farcs21(0\farcs93)$	&	26.5(4.2)	&	P241+40	\\
J1619$+$3953	&		&	1884	&	23	&	J161913.40$+$395306.2	&	$+0\farcs78(0\farcs61)$	&	$-0\farcs96(1\farcs17)$	&	1.5(0.5)	&	P245+40	\\
J1624$+$5850	&		&	651.8	&	26.4	&	J162400.98$+$585015.8	&	$-0\farcs11(0\farcs30)$	&	$-0\farcs09(0\farcs27)$	&	3.4(0.5)	&	P246+58	\\
J1628$+$4406	&	L/V	&	181.2	&	7	&	J162850.24$+$440642.5	&	$+0\farcs71(0\farcs25)$	&	$+0\farcs09(0\farcs25)$	&	12.4(1.9)	&	P247+45	\\
J1630$+$3550	&	L	&	3.229	&	17	&	J163035.93$+$355042.5	&	$+0\farcs19(0\farcs28)$	&	$-0\farcs06(0\farcs28)$	&	5.9(0.9)	&	P248+35	\\
J1630$+$3734	&	L	&	3.318	&	14	&	J163036.45$+$373441.9	&	$+0\farcs20(0\farcs27)$	&	$+0\farcs07(0\farcs26)$	&	4.3(0.7)	&	P249+38	\\
J1638$+$4005	&		&	767.7	&	33	&	J163816.24$+$400556.5	&	$+0\farcs09(0\farcs32)$	&	$-0\farcs15(0\farcs27)$	&	3.1(0.5)	&	P248+40	\\
B1639$+$36A	&		&	10.38	&	30	&	J164140.90$+$362714.2	&	$-0\farcs38(0\farcs86)$	&	$+0\farcs65(0\farcs71)$	&	1.1(0.3)	&	P249+38	\\
B1639$+$36B	&		&	3.528	&	29.45	&	J164140.37$+$362558.2	&	$+0\farcs32(0\farcs64)$	&	$+0\farcs29(0\farcs47)$	&	1.1(0.3)	&	P249+38	\\
J1647$+$6608	&	L/V	&	1600	&	22.55	&	J164732.46$+$660821.9	&	$+0\farcs35(0\farcs29)$	&	$+0\farcs25(0\farcs31)$	&	7.6(1.1)	&	P254+65	\\
J1656$+$6203	&		&	776.2	&	35.26	&	J165610.32$+$620350.6	&	$-0\farcs23(0\farcs41)$	&	$-0\farcs20(0\farcs39)$	&	1.6(0.3)	&	P253+63	\\
J1657$+$3304	&		&	1570	&	24	&	J165750.67$+$330434.0	&	$+0\farcs20(0\farcs31)$	&	$-0\farcs44(0\farcs30)$	&	3.1(0.5)	&	P253+33	\\
J1658$+$3630	&	L/V	&	33.03	&	3	&	J165826.55$+$363030.4	&	$-0\farcs33(0\farcs22)$	&	$-0\farcs37(0\farcs22)$	&	29.5(4.5)	&	P255+38	\\
J1707$+$3556	&		&	160	&	19	&	J170702.75$+$355636.5	&	$-0\farcs22(0\farcs29)$	&	$-0\farcs05(0\farcs29)$	&	4.4(0.7)	&	P255+35	\\
J1715$+$4603	&		&	548	&	21	&	J171543.76$+$460400.5	&	$+0\farcs12(0\farcs95)$	&	$-1\farcs35(1\farcs06)$	&	0.6(0.3)	&	P258+45	\\
J1717$+$4308A	&		&	3.16	&	35	&	J171706.53$+$430802.6	&	$-0\farcs40(0\farcs55)$	&	$+0\farcs85(1\farcs03)$	&	1.5(0.4)	&	P258+43	\\
J1722$+$3519	&	L/V	&	822	&	24	&	J172209.51$+$351918.7	&	$-0\farcs01(0\farcs24)$	&	$-0\farcs17(0\farcs24)$	&	13.9(2.1)	&	P261+35	\\
J1741$+$3855	&		&	829	&	47.22	&	J174112.34$+$385510.1	&	$-0\farcs03(0\farcs37)$	&	$-0\farcs25(0\farcs36)$	&	3.3(0.5)	&	P265+38	\\
J1745$+$4254	&		&	305	&	38	&	J174550.14$+$425438.0	&	$-0\farcs19(0\farcs37)$	&	$-0\farcs13(0\farcs35)$	&	2.5(0.4)	&	P265+43	\\
J1758$+$3030	&	V	&	947	&	35	&	J175825.86$+$303024.0	&	$+0\farcs27(0\farcs41)$	&	$-0\farcs08(0\farcs66)$	&	16.8(2.6)	&	P270+30	\\
B1811$+$40	&	L/V	&	931.1	&	41.56	&	J181313.20$+$401339.3	&	$+0\farcs88(0\farcs24)$	&	$-0\farcs26(0\farcs24)$	&	36.5(5.5)	&	P272+40	\\
J1816$+$4510	&	V	&	3.193	&	39	&	J181635.95$+$451033.9	&	$-0\farcs13(0\farcs21)$	&	$-0\farcs06(0\farcs21)$	&	76.0(11.4)	&	P273+45	\\
J1821$+$4147	&	V	&	1262	&	41	&	J182152.34$+$414702.6	&	$-0\farcs00(0\farcs27)$	&	$-0\farcs12(0\farcs25)$	&	6.4(1.0)	&	P276+43	\\
J2155$+$2813	&		&	1609	&	77	&	J215515.22$+$281309.3	&	$+7\farcs93(0\farcs33)$	&	$+2\farcs75(0\farcs36)$	&	4.7(0.7)	&	P330+28	\\
J2156$+$2618	&		&	498	&	48	&	J215623.71$+$261829.9	&	$-0\farcs14(0\farcs53)$	&	$+0\farcs54(0\farcs64)$	&	1.8(0.4)	&	P330+26	\\
J2202$+$2134	&		&	498	&	17.7	&	J220217.00$+$213434.0	&	$-0\farcs27(0\farcs75)$	&	$-0\farcs47(0\farcs58)$	&	4.5(0.8)	&	P331+21	\\
J2204$+$2700	&		&	84.7	&	35	&	J220443.71$+$270054.9	&	$-1\farcs28(0\farcs73)$	&	$-0\farcs31(0\farcs76)$	&	0.9(0.3)	&	P330+28	\\
J2212$+$2450	&	L/V	&	3.908	&	25.21	&	J221227.63$+$245036.7	&	$-0\farcs80(0\farcs32)$	&	$+0\farcs08(0\farcs29)$	&	8.3(1.3)	&	P333+26	\\
B2210$+$29	&	V	&	1005	&	75	&	J221223.34$+$293305.4	&	$+0\farcs07(0\farcs23)$	&	$-0\farcs11(0\farcs23)$	&	15.6(2.3)	&	P333+31	\\
J2214$+$3000	&	L	&	3.119	&	23	&	J221438.84$+$300038.2	&	$+0\farcs13(0\farcs29)$	&	$-0\farcs07(0\farcs28)$	&	4.7(0.7)	&	P333+31	\\
J2222$+$2923	&		&	281.4	&	49	&	J222303.22$+$292358.9	&	$+0\farcs05(0\farcs36)$	&	$-0\farcs32(0\farcs30)$	&	4.2(0.7)	&	P336+28	\\
J2227$+$3038	&	V	&	842	&	20	&	J222741.77$+$303820.8	&	$+0\farcs03(0\farcs38)$	&	$-0\farcs03(0\farcs30)$	&	17.0(2.6)	&	P336+31	\\
J2229$+$2643	&	L	&	2.978	&	23	&	J222950.89$+$264357.4	&	$-0\farcs17(0\farcs24)$	&	$+0\farcs15(0\farcs24)$	&	12.1(1.8)	&	P338+26	\\
J2234$+$2114	&	V	&	1359	&	35	&	J223456.63$+$211419.3	&	$+0\farcs18(0\farcs50)$	&	$-0\farcs55(0\farcs55)$	&	23.1(3.5)	&	P340+21	\\
B2303$+$30	&		&	1576	&	50	&	J230558.32$+$310001.2	&	$+0\farcs06(0\farcs22)$	&	$+0\farcs00(0\farcs22)$	&	30.3(4.6)	&	P345+31	\\
J2306$+$3124	&	L	&	342	&	46	&	J230619.22$+$312420.2	&	$-0\farcs25(0\farcs24)$	&	$+0\farcs15(0\farcs24)$	&	9.1(1.4)	&	P345+31	\\
B2315$+$21	&	L/V	&	1445	&	21	&	J231757.85$+$214947.1	&	$-0\farcs08(0\farcs22)$	&	$+0\farcs84(0\farcs22)$	&	56.6(8.6)	&	P348+21	\\
J2322$+$2057	&		&	4.808	&	13	&	J232222.35$+$205702.8	&	$-0\farcs23(0\farcs43)$	&	$-0\farcs18(0\farcs43)$	&	1.6(0.3)	&	P351+21	\\
J2350$+$3140	&		&	508	&	39	&	J235041.19$+$314047.4	&	$+0\farcs18(0\farcs26)$	&	$-0\farcs20(0\farcs26)$	&	5.3(0.8)	&	P357+31	\\
J2355$+$2246	&		&	1841	&	23	&	J235549.35$+$224610.6	&	$+6\farcs25(4\farcs51)$	&	$+6\farcs38(8\farcs01)$	&	3.0(0.5)	&	P000+23

\end{longtable}

\tablefoot{
Pulsars that we detect by cross-matching pulsars from PSRCAT to sources from LoTSS DR2. The left side of the table shows from left to right: the J- or B-name from pulsars, the letters L and/or V to indicate that a pulsar has been detected in linear polarization from \citet{O'Sullivan+2023} and/or circular polarization from \citet{V-LoTSS}, the spin-period ($\mathrm{P}$) of the respective pulsar in milliseconds and the Dispersion Measure ($\mathrm{DM}$) in pc/cc. The right side of the table shows from left to right: the ILTJ-name of LoTSS sources, the separation in R.A. ($\mathrm{\Delta \alpha}$), the separation in Dec. ($\mathrm{\Delta \delta}$), the flux density at 144\,MHz from LoTSS ($\mathrm{S_{144}}$) and the LoTSS field corresponding to the LoTSS counterpart of the pulsar. \\
\tablefoottext{a}{Reported by \citealt{1995ApJ...455L..55N} and \citet{TULLIP}, but not detected in \citet{O'Sullivan+2023}.}
\tablefoottext{b}{This source is an extended source made up of two point-sources (Fig.\,\ref{fig:pulsar_examples}f).}
\tablefoottext{c}{Reported by \citet{2015A&A...583A.137J}, but not detected in \citet{O'Sullivan+2023}.}
\tablefoottext{d}{This source is not in the LoTSS catalog (Fig.\,\ref{fig:pulsar_examples}i).}
\tablefoottext{e}{This source is not in the LoTSS catalog (Fig.\,\ref{fig:pulsar_examples}e).}
}

\end{appendix}

\twocolumn

\end{document}